\documentclass[aps,11pt,amsmath,amssymb,nofootinbib,a4paper,showpacs]{revtex4}
\usepackage[breaklinks]{hyperref}
\usepackage[latin1]{inputenc}
\usepackage{graphicx}
\usepackage{color}
\usepackage{amsfonts,amsthm}
\usepackage{bm,bbm}
\usepackage[OT2,OT1]{fontenc}
\newcommand\cyr{%
\renewcommand\rmdefault{wncyr}%
\renewcommand\sfdefault{wncyss}%
\renewcommand\encodingdefault{OT2}%
\normalfont
\selectfont}
\DeclareTextFontCommand{\textcyr}{\cyr}

\begin{document}
\title{The Bronstein hypercube of Quantum Gravity}
\author{{\bf Daniele Oriti}}
\affiliation{Max Planck Institute for Gravitational Physics (Albert Einstein Institute) \\ Am Muehlenberg 1, D-14476 Potsdam-Golm, Germany, EU}
\affiliation{II Institute for Theoretical Physics, University of Hamburg,  Luruper Chaussee 149, 22761 Hamburg, Germany, EU}
\email{daniele.oriti@aei.mpg.de}
\begin{abstract}
We argue for enlarging the traditional view of quantum gravity, based on \lq quantizing GR\rq,  to include explicitly the non-spatiotemporal nature of the fundamental building blocks suggested by several modern quantum gravity approaches (and some semi-classical arguments), and to focus more on the issue of the emergence of continuum spacetime and geometry from their collective dynamics. We also discuss some recents developments in quantum gravity research, aiming at realising these ideas, in the context of group field theory, random tensor models, simplicial quantum gravity, loop quantum gravity, spin foam models.

\end{abstract}
\maketitle
\section{Introduction}
The quest for quantum gravity has undergone a dramatic shift in focus and direction in recent years. This shift followed, and at the same time inspired and directly produced many important results, further supporting the new perspective. The purpose of this contribution is to outline this new perspective, and to clarify the conceptual framework in which quantum gravity should then be understood. We will emphasize how it differs from the traditional view and new issues that it gives rise to, and we will frame within it some recent research lines in quantum gravity. 

Both the traditional and new perspectives on quantum gravity are nicely captured in terms of a \lq diagram in the space of theoretical frameworks\rq. The traditional view can be outlined in correspondence with the \lq Bronstein cube\rq ~of physical theories \cite{Bronstein}. The more modern perspective, we argue, is both a deepening of this traditional view, and a broader framework, which we will outline using (somewhat light-heartedly) a \lq Bronstein hypercube\rq ~of physical theories.

\section{The Bronstein cube of quantum gravity}
The Bronstein cube of quantum gravity \cite{Bronstein} is FIG.1. It lives in the $cGh$ space, identified by the three axes labeled by Newton's gravitational constant $G$, the (constant) velocity of light $c$, or, better, its inverse $1/c$, and Planck's constant $h$. Its exact dimensions do not matter, the axes all run from $0$ to infinity, but its corners can be identified with the finite values that the same constants take in modern physical theories. 

\begin{figure}[h]\label{fig:Bronsteincubefig}
\includegraphics[scale=0.3]{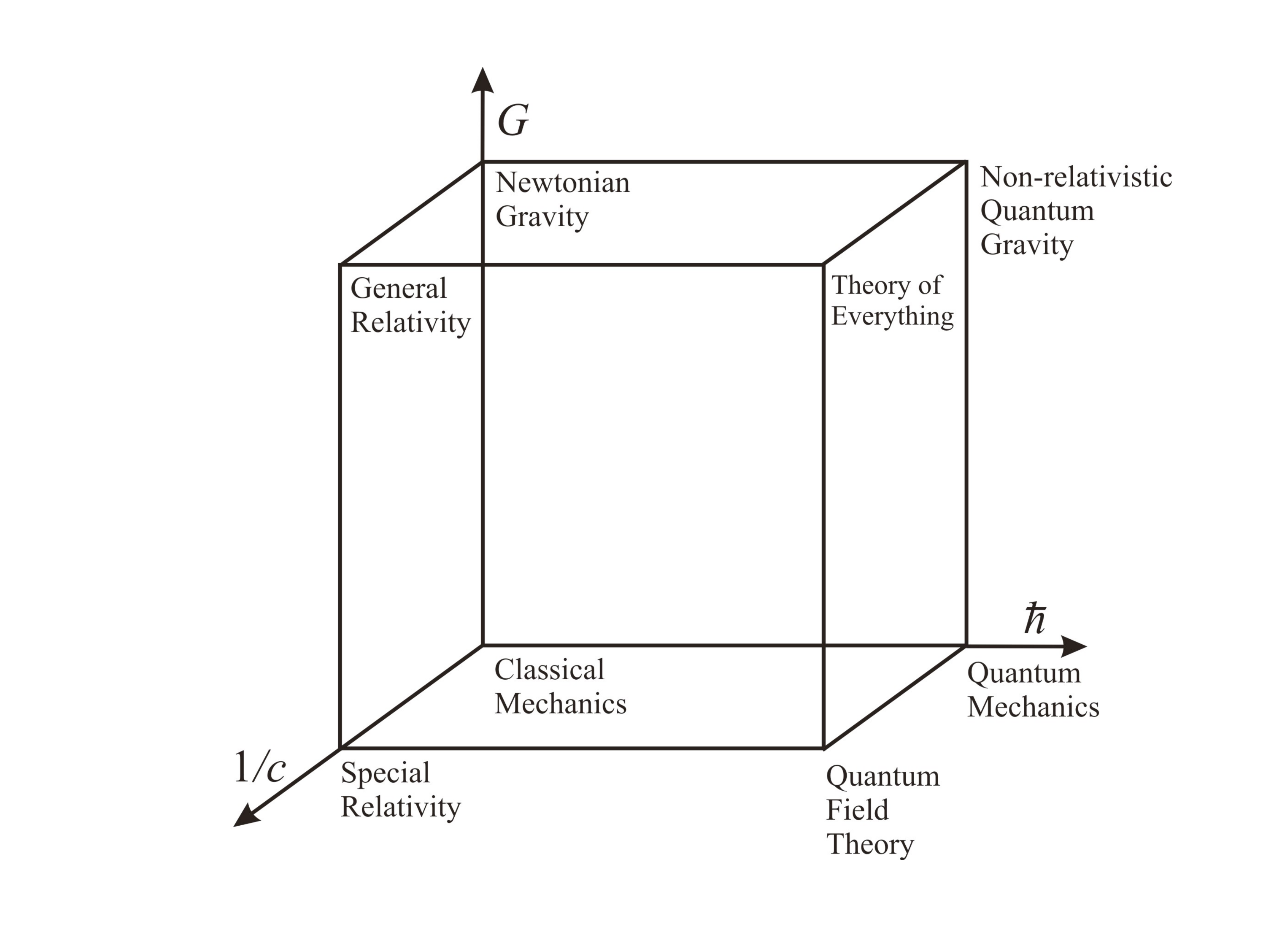}
\caption{The Bronstein cube; the picture is taken, with many thanks to the author, from S. Hossenfelder's blog at http://backreaction.blogspot.com/2011/05/cube-of-physical-theories.html.}
\end{figure}

The picture does not represent specific physical theories or models (despite some of the names used in the same picture), but more general theoretical {\it frameworks}.
Its conceptual meaning can be understood by moving along its corners, starting from the simplest theoretical framework, i.e. classical mechanics, located at the origin $(0,0,0)$ (understood as hosting all theories and models formalised within this framework, be them about fields, particles, forces). Moving from the origin along the $G$-axis, we start including in our theoretical framework gravitational physics, i.e. the effects of the gravitational  interactions on the same entities dealt with in classical mechanical models. The very moment these become non-zero, we are in the realm of classical Newtonian gravity. If we move instead from the origin along the $1/c$ axis, we start taking into account relativistic effects, i.e. due to the finite propagation speed of physical signals (information), bodies and interactions, bounded by the velocity of light.  If both relativistic effects and gravitational ones are taken into account, we reach the corner presided by General Relativity, a classical, relativistic mechanics including also the gravitational interactions of all mechanical systems. Historically, this is the corner reached with the first revolution of 20th century physics. The other revolution came with the realization that an altogether different \lq direction\rq ~exists, in the physical world: quantum phenomena, those (roughly) due to the existence of a finite lower bound for the \lq action\rq ~of a system, for the area it can occupy in phase space, corresponding to the Planck constant $h$. Thus, moving from the origin along the $h$-direction, we find quantum mechanics, the modern framework for all physical systems, with its associated amount of weirdness and marvels, which we have not yet grown fully accustomed to. Actually, it becomes the modern framework for all physical systems once we take into account also relativistic features, by moving also along away from the non-relativistic side of the cube, entering the domain of quantum field theory. This is indeed the modern framework of physics. Or is it? Not really, of course, since we know that all systems are quantum and relativistic, but we also know that gravity exists and that, in the more modern understanding coming from GR, the gravitational field, the spacetime geometry that is identified with it, and thus spacetime itself, is a physical, dynamical entity. Modern physics is somehow framed either close to the GR corner, or around the QFT corner, but it cannot be said to correspond to any single domain within the Bronstein cube, lacking a quantum theory of gravity. We would then like to be able to move along both the $h$-direction and the $G$-direction, incorporating both gravitational effects (including very strong ones) and quantum effects into a single coherent description of the world. The corner we would reach by constructing a quantum gravity theory would be that of a \lq theory of everything\rq , not in the sense of any ontological unification of all physical systems into a single physical entity (although that is a possibility, and a legitimate aspiration for many theoretical physicists), but simply in the sense that in such framework we could in principle describe in a formally unified way all known types of phenomena: quantum, relativistic, gravitational.

Obviously, the above is but an extremely rough sketch of theoretical physics. It does not account even remotely for the complexity of phenomena that are actually described by the mentioned frameworks. And it does not say anything about the very many subtleties involved in actually moving from one framework to the other, and back from there.  One example should suffice to illustrate these limitations: the classical limit, which naively should allow us to reduce a quantum (description of a) system to its classical counterpart. Taking this limit and understanding how the classical world emerges from the quantum one is a notoriously thorny topic, involving mathematical complications and conceptual ones, including the issue of measurement, decoherence, etc.
Other limitations have to do with the fact that all the quantities appearing in the picture are dimensionful, and thus they do not correspond to directly observable/measurable quantities. Plus, no mention is made of the actual nature of the systems considered, which however modifies greatly what can actually be described (and how) within each framework. For example, we know that a relativistic description of quantum interacting particles is problematic and requires to move to a field-theoretic framework, being then understood only as an excitation of a quantum field. Thus, not every entity can live in every corner of the Bronstein cube.
  
Before we discuss how the Bronstein cube encodes the problem of quantum gravity, and what is missing about it from a more modern standpoint, it is useful to point out what else is missing in it, from the landscape of modern theoretical physics. Entirely missing  is the whole realm of statistical mechanics and condensed matter theory, even though this can be argued to represent the third revolution in modern physical thinking occurred during the last century. The list of new phenomena and surprising lessons about the natural world that this third revolution has brought to us is very long indeed: from the whole idea of the renormalization group, to the rich theory and phenomenology of phase transitions, including the discovery of new phases of matter; from superfluidity and superconductivity to new materials; from the ubiquitous role of symmetry and symmetry breaking to the role of quantum information theory in understanding quantum many-body systems.  One argument against considering all of this on the same footing as the theoretical frameworks occupying the corners of the Bronstein cube could be that they are not \lq fundamental physics\rq , since  they are all complicated results of simple laws corresponding to the quantum mechanics of many atoms or electrons interacting mostly via the electromagnetic field. This objection, however, rests on a rather rigid reductionist attitude that has been convincingly challenged in the past \cite{emergence} and we find untenable. One can consider condensed matter theory not fundamental when looking at purely ontological aspects, since it deals with \lq derived entities\rq ~or \lq derived laws\rq ~for fundamental entities (and even this could be challenged, since it assumes a very rigid ontology), but the Bronstein cube, as we presented it, is not about the ontology of the world. From a more epistemological perspective, on the other hand, the discovery of emergent features from collective behaviour of fundamental entities, as well as the very complexity, richness and stability of this emergent realm of phenomena is a fundamental change in our view of the world, and should be considered as radical and important as those corresponding to the three axes of the Bronstein cube (see again the various contributions in \cite{emergence}). This is not a new idea. For example, it had been proposed to add the Boltzmann constant to the other three, and extend accordingly the theory space as pictured in the Bronstein cube \cite{Cohen-Tannoudji}. This is not the extension that we will discuss in the following, as ours will be directly motivated by quantum gravity considerations.  However, we will see that our extension can also be seen as incorporating the insights of this third physics revolution.   

\section{The problem of quantum gravity from the Bronstein cube}
Before thinking about such an extension, let us dig more into the perspective on quantum gravity as encoded in the Bronstein cube.  This is the straightforward view that sees quantum gravity as obtained from quantizing the geometry (metric/gravitational field) of spacetime and its dynamics, by whatever quantization method, i.e. as \lq quantized GR\rq. The quantum gravity corner can be reached by incorporating quantum effects starting from the GR corner, or by adding the gravitational aspects of the world (equivalently, non-trivial spacetime geometries) in a quantum description of it. This corresponds to the strategy of all the traditional approaches to quantum gravity \cite{kiefer}: canonical quantum gravity, including, at least in its original form, the connection-based version of this programme corresponding to loop quantum gravity \cite{LQG}; covariant path integral formalisms, including discretized versions of the same like quantum Regge calculus \cite{qRC} and (causal) dynamical triangulations \cite{CDT}, to the extent in which the lattice structures  are understood only as regularization tools. It also includes the asymptotic safety programme \cite{AS}, based on the non-perturbative completion of the formulation of (perturbatively quantized) gravity as an effective field theory. It could include also (depending on the interpretation) the non-commutative geometry programme \cite{NCG}, where one quantizes the geometric structures of spacetime directly, without relying on their role as encoding the gravitational interaction. The situation with string theory \cite{ST} is more ambiguous, due to the huge variety of formalisms and research directions now under such umbrella label; still, the understanding of string theory as quantized GR may at least apply to its very early perturbative versions, since the interpretation as quantum gravity theories was due to the existence, in their spectrum, of graviton excitations, quanta of the gravitational field from an effective quantum field theory point of view. If string theory is used in this perturbative, semi-classical form to study the gravitational side of the AdS/CFT correspondence (which in itself may not require string theory at all), then also the latter would not present immediately a challenge to the usual picture of quantum gravity. Still, as we will discuss in the following, the AdS/CFT correspondence itself can be seen as a reason to believe that string theory itself requires a more radical departure from the conventional understanding of spacetime and geometry.

This perspective makes perfect sense  and exhausts the range of possibilities if the step from classical to quantum gravity does not entail a change of fundamental degrees of freedom, i.e. if the spacetime geometry, the metric field, gravity are primary entities and the task is to understand their quantum properties. Even in this case, of course, understanding their quantum properties may reveal a number of surprising and very exotic aspects of the world. The step from classical to quantum, when dealing with such a fundamental entity like spacetime geometry, is by all means a challenging one, both mathematically and conceptually. The issue of time, the debate between relationalism versus substantivalism in spacetime theories, the problems with diffeomorphism invariance, on top of the purely technical issues faced by quantum gravity theorists are there to testify the magnitude of the challenge. These issues are already challenging in a classical GR context, where the theory is complete and the physics is well understood. In a \lq quantized GR\rq ~context, what the theory is {\it expected} to involve, even leaving aside its incomplete status, raises a host of new and even more severe difficulties \cite{conceptual-kiefer}. One example will suffice:  what is left of usual physics, of the customary understanding of the world, in a theory with indefinite and fluctuating causal structures, an immediate consequence of superposing quantized geometries, even assuming that each of them maintains a continuum and close-to-classical character? This explains the difficulties in constructing a theory of quantum gravity on such basis, despite the many results obtained over a span of decades.  

\section{Beyond the Bronstein cube: the idea of emergent spacetime}
The point is, however, that the above perspective does not capture the range of problems faced by modern approaches to quantum gravity, neither at the technical not at the conceptual level. It does not capture where quantum gravity stands with respect to the rest of fundamental physics either. Let us explain why by first reviewing briefly a number of hints challenging the above view. They were produced by research {\it within} the Bronstein cube, but at the same time pushing against its walls, so to speak, noticing their fictitious nature and thus suggesting strongly that there is more to be done and discovered outside it.

These recent results are of two different types. First, they come from research directions not directly aiming at constructing a full theory of quantum gravity, but focusing on semi-classical gravitational physics, and sometimes on systems that are not gravitational at all but which give surprising insights on the possible nature of geometry and gravity. Second, they come directly from quantum gravity approaches, often of the \lq conservative\rq ~tradition living inside the Bronstein cube, which nevertheless end up producing challenges to the very perspective that inspired them.

The first group of results can be taken as suggesting that the continuum geometric structure of spacetime, on which General Relativity and quantum field theory are based, is not fundamental and that some sort of discrete quantum counterpart should replace it in a full theory of quantum gravity. If this is the case, spacetime as we know it would be an approximate, emergent notion from something else, which would be then not spatiotemporal in the usual sense (although it may retain some features of the spacetime we are accustomed to). The key points here are the discrete nature of the more fundamental degrees of freedom, and the need to see spacetime (and its geometry) as emergent.

The second group of results offers a number of proposals for what the more fundamental degrees of freedom could look like and for which features of continuum spacetime and geometry could be dropped in the more fundamental description. They also explore a range of more or less radical departures from \lq accepted behaviour\rq, that we may need to be accustomed to if we want to understand the more fundamental nature of space and time (and how to understand the world in their absence). 
 
Among the first group, the oldest results can be taken to be those establishing the existence of singularities in gravitational physics. They may be taken to imply only that quantum corrections to gravitational dynamics have to be taken into account, but they may also be taken as a suggestion that something more radical happens: a breakdown of the continuum spacetime description itself. The divergences of quantum field theory too admit a conservative as well as a more radical interpretation. The correct coupling of quantum field theories for matter (and other interactions) with a properly quantized version of GR may be all that is needed to cure them, introducing a natural cutoff scale. Or they may be an indication that some more radical form of discreteness replaces the continuum nature of quantum fields, including the gravitational field. Indeed, several scenarios incorporating a minimal length (or a maximal energy scale) have been proposed \cite{sabine}, as effective descriptions of quantum gravity, and they end up challenging many more aspects of standard spacetime physics, including for example locality, which is at the root of quantum field theory and of the whole of continuum spacetime physics. This challenge to locality is not surprising, since the hypothesis of a minimal length was proposed from the very beginning as a consequence of the impossibility of exact localization when the gravitational effects of quantum measurements are taken into account. Among these scenarios, many rely on non-commutative geometry tools \cite{NCG}. Thus, they also offer a first example of a quantum gravity approach that can be understood at first in a conservative way, and that turns out to be more radical than imagined in its implications. To this group belong also the very many results dealing with black hole thermodynamics and in particular with black hole entropy \cite{BHthermo}. These are far too many (and interesting) to review them here. However, the main message, for our concerns, is simple.  A black hole, in the end, is a region of spacetime. If it has entropy, and it is a standard statistical mechanical system, then this entropy accounts for its microstates. These microstates could be (a subset of) modes of the quantized metric field or of some relevant matter field (inside or close to the horizon). But they could also be microstates associated to a different set of dynamical entities that only at macroscopic scales look like any of the two.
 Moreover, if this entropy is finite, this microstructure should have some in-built fundamental discreteness, and thus be of a very different nature than ordinary spacetime (and geometry). Similarly radical are the results that support an holographic nature for the degrees of freedom constituting a black hole, and that stem from a combination of classical and semi-classical arguments. In turn, semi-classical black hole physics has inspired a number of research directions investigating the more general thermodynamical properties of spacetime and the possibility that spacetime/gravitational dynamics (including the whole of GR) is itself to be understood as the thermodynamics, or hydrodynamics, in some of the approaches, of unknown microscopic degrees of freedom \cite{ST-thermo}. In this view, the spacetime metric, thus the gravitational field, would be a coarse-grained variable accounting for such microscopic degrees of freedom, and spacetime should be understood as a sort of a fluid-like collective entity. Another independent research area makes implicitly the same suggestion: analogue gravity models in condensed matter systems \cite{analogue}, in particular in the context of quantum fluids. These systems reproduce, at the hydrodynamic level, several phenomena with an equivalent description in terms of semi-classical physics on a curved geometry, including semi-classical black hole-like physics. Thus they support the suggestion that the gravitational physics is an effective, emergent description of a different type of physics, and spacetime indeed a fluid-like system, the result of the collective behaviour of non-spatiotemporal entities.

Quantum many-body systems have also brought us a different type of surprise, showing an intriguing connection between entanglement and geometry, suggesting that the latter can be reduced to the former instead of being treated as fundamental \cite{entanglement-geometry}. For example: the entanglement entropy associated to a region A on the flat boundary of an AdS space, computed within a simple CFT, is proportional to the area of the minimal surface inside the bulk AdS space with the same boundary as A;  the mutual information between two spatial regions on the same flat boundary scales inversely with the geodesic distance between the two regions, measured again in the bulk AdS;  the very connectivity between two regions of spacetime has been conjectured to be due to the entanglement between (the quantum degrees of freedom of) the two regions. 

Many of these results have been obtained in the context of the AdS/CFT correspondence \cite{AdS/CFT}. This can be seen as an approach to quantum gravity (at least the sector of it corresponding to AdS boundary conditions), that, despite relying so far mostly on standard field theory methods, suggest a more radical view of spacetime and gravity (understood as curved geometry), in which the latter is again emergent from a system which, while defined on a continuum flat spacetime, is not gravitational. AdS/CFT is then also a first example of the second group of results, pointing to a view of the quantum gravity problem beyond the Bronstein cube (for a discussion on the conceptual challenges raised by the AdS/CFT correspondence, see \cite{AdS/CFT-phil}).

String theory \cite{ST} is often used to describe the gravitational side of the AdS/CFT correspondence, and it has been another independent source of radical challenges to the conventional view of spacetime and of quantum gravity. These range from the implications of T-duality for the notion of spatial distance itself, to the equivalence between different spacetime topologies encoded in mirror symmetry, to the generalised geometries that seem to be needed to describe various effective configurations of string theories \cite{gengeoST}. The upshot is that, while we do not know what sort of fundamental degrees of freedom underlie string theories, we know that they will not be spatiotemporal or geometric in any standard sense, and that spacetime and geometry as we know them are both collective and emergent notions, in such context \cite{seiberg}.   
  
Generalised geometries, in particular fractal geometries \cite{gianluca}, have been studied extensively, also because the running of spacetime dimensions found in several quantum gravity approaches suggest a role for them in the full theory, maybe as an intermediate regime between the fundamental,  non-spatiotemporal one and the emergent spacetime of standard field theory. 

In fact, quantum gravity approaches, even when starting as conservative quantizations of GR, ended up proposing concrete candidates for the fundamental degrees of freedom underlying spacetime which are, on their own, not spatiotemporal in the standard sense. Loop quantum gravity \cite{LQG} has fundamental quantum states encoded in spin networks, graphs labeled by group representations, with histories corresponding to cellular complexes labeled by the same algebraic data (spin foams) \cite{SF}. The same type of quantum states (and discrete histories) are shared with group field theories \cite{GFT}. These states and histories, in appropriate regimes, can be put in correspondence with piecewise-flat (thus discrete, and singular) geometries, but in the most general cases they will not admit even such proto-geometric interpretation. The latter are, in turn, the building blocks of simplicial quantum gravity approaches like quantum Regge calculus \cite{qRC} and (causal) dynamical triangulations \cite{CDT}, which indeed can be seen both as strictly related to group field theories and their purely combinatorial counterparts, random tensor models \cite{TM}. In all these quantum gravity formalisms, therefore, continuum spacetime has to emerge from structures which are fundamentally discrete and rather singular (from the continuum geometric perspective), and in some cases, purely combinatorial and algebraic. A different type of fundamental discreteness, not less radical, is the starting point of the causal set approach \cite{CS}. 

The main lesson seems to be that continuum spacetime and geometry have to be replaced, at the fundamental level, by some sort of discrete, quantum, non-spatiotemporal structures, and have to emerge from them from their collective dynamics, in some approximation \cite{oriti-emergence}.

One could say that, since these \lq atoms of space\rq  ~are assumed to be quantum entities, spacetime is understood in all these formalisms as a peculiar quantum many-body system \cite{oriti-manybody}, that only at macroscopic scales will look like the smooth (indeed, fluid-like) object we are accustomed to. There is, therefore, an obvious general coherence between this picture of spacetime painted by quantum gravity approaches, and the more indirect (but also more closely related to established physics) indications obtained by the semi-classical considerations, e.g. in black hole physics.

\section{The Bronstein hypercube of quantum gravity}
This has one general consequence for our understanding of the quantum gravity problem. Given such fundamental (non-geometric, non-spatiotemporal) degrees of freedom, there is one new direction to explore: from small to large numbers of such fundamental entities. We know (from quantum many-body systems and condensed matter theory) that the physics of few degrees of freedom is very different from that of many of them. When taking into account more and more of the fundamental entities and their interactions, we should expect new collective phenomena, new collective variables more appropriate to capture those phenomena, new symmetries and symmetry breaking patterns, etc. And it is in the regime corresponding to many fundamental building blocks that we expect a continuum geometric picture of spacetime to emerge, so that the usual continuum field theory framework for gravity and other fields will be a good approximation of the underlying non-spatiotemporal physics.  

Notice that the above is to a large extent independent of whether the discrete structures are understood as physical entities or simply as regularization tools. This will affect, of course, whether one assigns a physical interpretation to all the results of their collective behaviour or not, and whether or not one tries to eliminate any signature of the discrete structures leading to them. But the existence of the mentioned new direction remains a fact, as it remains true that one has to learn to move along this new direction, if one wants to recover a continuum picture for spacetime and geometry (and with them, a gravitational field with relativistic dynamics). 

To have a better pictorial representation of what quantum gravity is about, then, the Bronstein cube should be extended to an object with four (a priori) independent directions, to a \lq Bronstein hypercube\rq, as in FIG.2.

\begin{figure}[h]\label{fig:hypercubefig}
\includegraphics[scale=0.35]{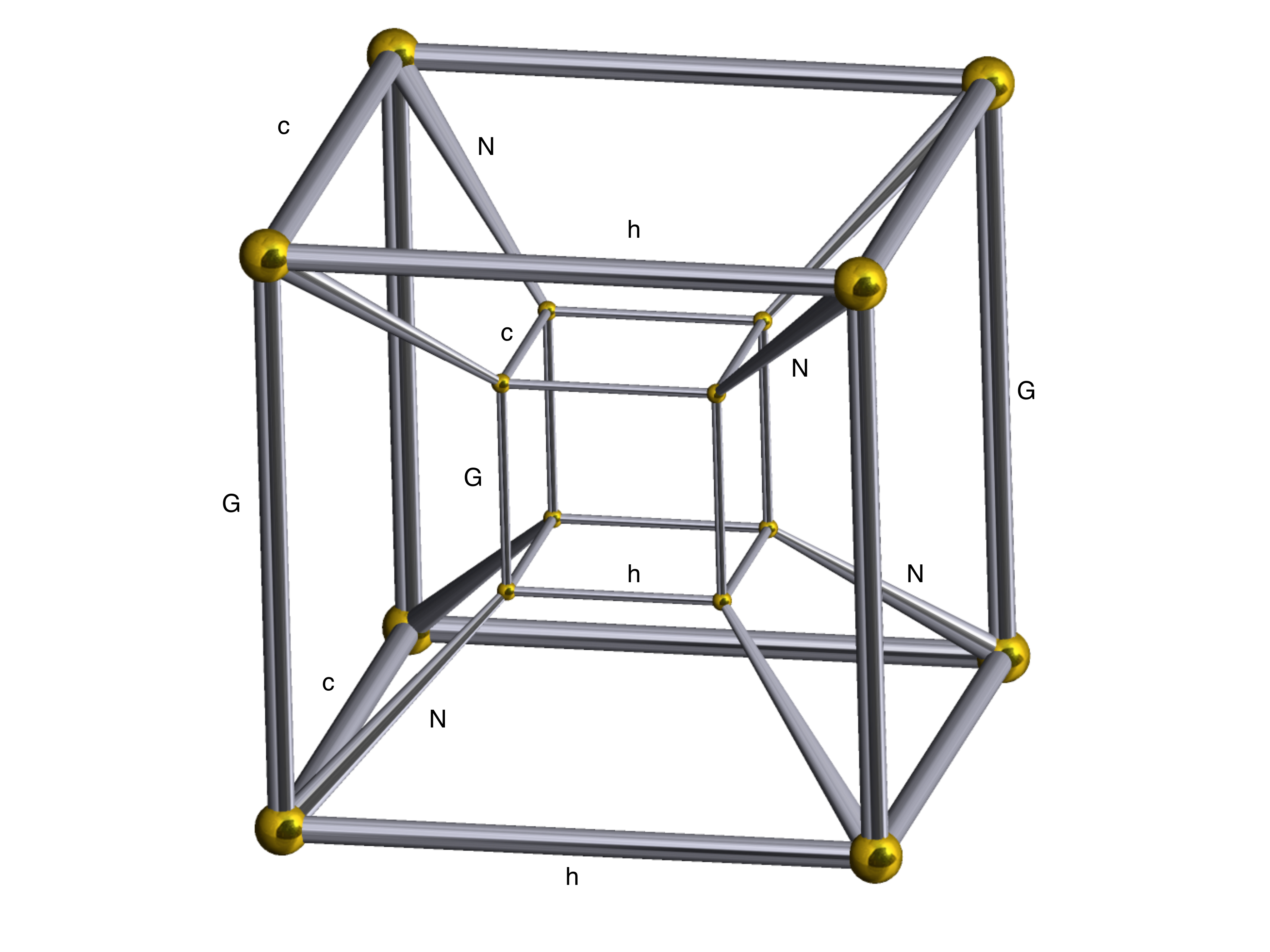}
\caption{The Bronstein hypercube. The picture is an adapted version of the hypercube at: https://en.wikipedia.org/wiki/Tesseract}
\end{figure}

The fourth direction is labeled $N$, to indicate the number of quantum gravity degrees of freedom that need to be controlled to progressively pass from an entirely non-geometric and non-spatiotemporal description of the theory to one in which spacetime can be used as the basis of our physics. A complete theory of quantum gravity will sit at the same corner in which it was sitting in the Bronstein cube (which is obviously a subspace of this hypercube), but the same theory admits a partial, approximate formulation at any point along the N-direction ending at that corner. Only, the more one moves away from it, the less the notions of continuum spacetime and geometry will fit the corresponding physics. 
One could say that the definition of a theory of quantum gravity will be provided in the opposite corner (looking only at the two ends of the N direction, while keeping both $G$, $h$ and $1/c$ finite), because it is at this point that the definition of the fundamental degrees of freedom of the theory and of their basic quantum dynamics will be put on the table. This is sensible, but it is also true that  providing a complete definition of the same theory amounts to making sure it is well-defined up to the opposite end, even though the same theory will always be used in some approximation or truncation.  

As we had anticipated, proposing the Bronstein hypercube as the proper arena for quantum gravity means stating that one needs to brings in the lessons and tools of statistical mechanics and condensed matter theory, i.e. the third revolution of last century's physics. It is in that context that we have learnt to control the rich physics of many quantum interacting degrees of freedom. 
We could label then the new direction of the Bronstein hypercube by the Boltzmann's constant, also in order to emphasize the above point \cite{Cohen-Tannoudji}. It could also be a way to make a link with information theory (another crucial area of developments in modern physics), with the implicit link between number of degrees of freedom of a system and its (Boltzmann) entropy, in turn hinting at the physical nature of such information content. This relabelling would have the advantage of characterizing the hypercubic extension of the Bronstein cube by the addition of a fourth fundamental constant, in many ways on equal footing as the other three. It is indeed useful to think in these terms. We do not use this relabelling explicitly simply because we want to maintain the focus on the number of (quantum gravity) degrees of freedom to be controlled in different regimes of the theory, rather than with any specific context in which the new degrees of freedom manifest their physical nature.
Another reason for not adopting a terminology directly reminiscent of statistical concepts is the following. We should not confuse the task of \lq moving along the $N$ direction\rq , that is, of understanding the continuum limit of the fundamental degrees of freedom of quantum gravity, and the emergence of continuum spacetime in the process, with the distinct issue, albeit related and definitely important, of defining a general relativistic (quantum) statistical mechanics, including the gravitational field and its thermal fluctuations \cite{carloStat}. Understanding the continuum physics of quantum gravity degrees of freedom may involve formulating a proper statistical framework for them, and this framework would have to be covariant in the obvious sense of not depending on any preferred spatiotemporal frame, local or global, for the simple reason that they would not be defined in any spatiotemporal context to start with. The key distinction has to do with the nature of the entities whose statistical framework one is considering. These fundamental quantum gravity degrees of freedom are not smooth spacetimes (or geometries), nor their straightforward quantised version. Thus, it is the difference between the (quantum) statistical treatment of the gravitational field (with coupled matter fields) and the (quantum) statistical treatment of \lq non-spatiotemporal building blocks of spacetime\rq (together with \lq non-spatiotemporal\rq ~building blocks of matter\rq) \cite{isha-KMS}. 

One obvious limitation of the description in terms of the \lq number of fundamental degrees of freedom\rq ~$N$ is that this notion is ambiguous. Not only what counts as \lq fundamental quantum gravity degrees of freedom\rq ~depends inevitably on the quantum gravity formalism under consideration, but the very notion of relevant degrees of freedom of a system is intrinsically ambiguous. For any quantum system, in fact, it depends on the vacuum state chosen (on the chosen irreducible representation of the fundamental algebra of its quantum observables) and on the adopted scale of description and on the observables chosen as relevant for capturing the physics one is interested in. The first aspect shows that the starting point adopted for describing the system is not god-given and one should keep this ambiguity under check; but one starting point needs to be chosen and once this is done, as in all quantum gravity approaches we know, our arguments about the need to \lq move along the $N$ direction\rq ~are valid. These ambiguities imply that there is no single hypercube of physical theories, because what one finds in its corners and along its edges depends not only on the specific quantum gravity framework being chosen, but also on specific criteria used to choose a set of degrees of freedom, a certain vacuum state, relevant symmetries, and so on, {\it within} each specific quantum gravity framework. The choice of all these elements is complex and leads, in general, to inequivalent theories. Once more, the Bronstein hypercube, just like the Bronstein cube, should be understood only as a representation of the {\it conceptual environment} within which specific, candidate physical theories of quantum gravity are to be studied and interpreted.
 The second aspect is a good part of the difficulties that we have to solve when moving along the $N$ direction. The number of fundamental degrees of freedom is a good proxy, in quantum gravity, for the usual notion of \lq scale\rq ~in usual spacetime physics, and the need for a change in description at different \lq scales\rq ~is exactly an important part of the notion of \lq emergence\rq, including the problem of the \lq emergence of continuum spacetime\rq. 

\section{Understanding the Bronstein hypercube}
The first point that using the Bronstein hypercube allows us to emphasize is the crucial distinction between classical and continuum limits. This is where the novelty of the new perspective on quantum gravity is most apparent. In a theory of quantum gravity in which the fundamental entity remains the gravitational field or the geometry of continuum spacetime, the problem of recovering the usual physics of GR (coupled to quantized fields) is the problem of controlling the classical approximation of the more fundamental quantum description of the same. If the fundamental discrete and quantum entities are not directly spatiotemporal,  the usual spacetime physics should emerge after taking some sort of continuum approximation, which may or may not be taken in conjunction with a semi-classical approximation. In fact, we have many examples in physics in which such continuum limit must be taken while maintaining the quantum properties of the fundamental constituents, and in which doing otherwise misses entirely important macroscopic physics.

Let us give some examples.
Consider some non-relativistic many-body quantum system of interacting atoms in flat space and with gravitational interaction switched off, i.e. the side of the Bronstein hypercube corresponding to $c>>1$, 
$G\sim 0$.  This definition is of course so generic that it includes an infinity of systems; in practice, all of condensed matter and solid state physics, and more. In the corner corresponding to small number of quantum atoms we have a bunch of discrete quantum entities and we can take two directions out of it. If we neglect their quantum properties (going towards the $h\sim 0$ area of the Bronstein hypercube), we obtain the classical mechanics of a few (point) particles. If we now take a continuum approximation by increasing the number of particles to infinity, we obtain a continuum classical system that, e.g. in the case of a fluid, could be described by classical hydrodynamics. 
Starting from the same corner but taking instead a continuum/hydrodynamic limit first, we could end up with a very peculiar continuum system like a quantum fluid, for example a Bose condensate (if we were dealing with spinless atoms at low temperatures), characterized by peculiar but very much physical features like superfluidity (or superconductivity) even at the macroscopic level. Or we could end up with even more exotic macroscopic phenomena, as in new phases of matter with topological order. All these macroscopic physical properties would be invisible if we were to take the continuum approximation after taking the classical approximation, or by taking the two simultaneously. Moreover, the two directions may be, in general, \lq non-commuting\rq ,  in the sense that taking the same two approximations in different order may give different results. The distinction between the two is therefore crucial.
A similar story can be told, starting with the same system in the same corner of the Bronstein hypercube, in terms of its statistical treatment, noticing the differences between classical and quantum statistical mechanics. And a similar picture can be drawn in the relativistic case, thus looking at the opposite side of the Bronstein hypercube, corresponding to $c\sim1$.

The distinction between the two independent directions of the Bronstein hypercube corresponding to $h$ and to $N$ is crucial also because they are travelled using very different types of mathematical techniques and of conceptual tools, and because we encounter very different types of new physics along the two paths. It is even more crucial in quantum gravity, for two main reasons. First, in the context of a theory so much under construction and so incomplete, and with so little guidance from empirical observations, it is very dangerous to not pay enough attention to the path we are taking or to have the wrong expectations about what we are supposed to do to make progress or about what phenomena we should look for at each stage. Second, in the case of quantum gravity we have really no obvious reason to expect that the two directions commute. 

The specifics depend of course on the quantum gravity formalism being considered. 
As an example, one can consider the relativistic and gravitational counterpart of the above atomic case, thus with $c\sim 1$ and $G\sim 1$, within the context of loop quantum gravity and/or group field theories. The corner with few quantum gravity degrees of freedom in the full quantum regime corresponds to a description of the world in terms of simple (superpositions of) spin networks associated to graphs with a smallish number of nodes, and with a quantum dynamics captured by amplitudes (spin foam models or lattice gravity path integrals) associated to (superpositions of) cellular complexes with limited combinatorial complexity (this could be increased, of course, but this description implies that it remains limited enough that we do not need to use different, collective or coarse grained entities and variables). The straightforward classical approximation of the same structures results in a description  in terms of classical piecewise-flat geometries (characterized, in the case of spin networks, by first order classical variables: edge lengths or triangle areas, and discrete connection variables), and with a dynamics encoded in solutions of discrete geometric equations, e.g. Regge geometries, possibly coming from the discretization of some continuum gravity action. Better, this is what we expect given the partial results we have so far \cite{SNclass, SFasymp}, but it is not completely established, and it remains an interesting challenge for the community. A further continuum limit is then needed to obtain an effective description in terms of (some possibly modified version of) GR and matter field theories. When the correspondence with Regge geometries (or similar) is solid, one can rely on the results obtained in that context for studying such continuum limit. When this is done, we are back in the sector of the Bronstein hypercube corresponding to the Bronstein cube, and we could consider quantizing our continuum gravitational theory. This already shows the difference between classical and continuum approximations in this quantum gravity context. Seen from the perspective encoded in the Bronstein hypercube, and given the same starting point, i.e. the same fundamental structures, the route towards a complete quantum theory of gravity would instead require taking into account more and more spin network degrees of freedom, at the quantum dynamical level, reaching (at least formally) the opposite corner of the hypercube along the N direction. This would be the regime of very large (possibly infinite) superpositions of quantum spin network states, including very refined graphs, with correspondingly complex interaction processes, which would then be effectively described in terms of collective variables like continuum fields (including the gravitational field) and field theoretic dynamics, e.g. some modified version of GR, up to any additional quantum corrections.
As we have emphasized above, we should not expect that the result will be the same that would be obtained by quantizing GR, and we should actually expect that this is not the case, if not under additional assumptions or approximations. Why this is the case should be clearer in the following, once we give a deeper look at what it is involved in moving along the $N$ direction.

\section{How to move along the $N$ direction} 
The $N$ direction is the path along which emergence of spacetime and geometry should take place. The notion of emergence itself  is a thorny topic in philosophy of science \cite{emergence-physics-phil}. The possibility that spacetime is not a fundamental but an emergent notion, and that the emergence process should then be understood in a non-spetiotemporal manner, raises a host of conceptual puzzles at both ontological and epistemological level. Some of them have been discussed in earlier work \cite{oriti-emergence}. Here we want to focus more on the physical aspect of this idea. That is, we want to discuss how we move along the $N$-direction, technically, and what we could expect to find, when we do so. 

So, first, how do we move along the $N$-direction, from less to more degrees of freedom? One main technical tool is the renormalization group, which is usually phrased as mapping a theory seen at a given scale to its counterpart at a different scale (this requires us to go beyond the naive view of renormalization as a way to \lq cure\rq ~or \lq hide\rq ~the theory's divergences). It is accompanied by several approximation schemes, by which we can extract suitable descriptions of the theory, capturing the key observables we are interested in, at the given scale.  As mentioned earlier, the notion of scale in the usual spacetime physics, e.g. in ordinary quantum field theory, intertwines the number of degrees of freedom of the system with some geometric or spatiotemporal quantity, like energy or distance, simply because such theories deal with degrees of freedom which are localised in spacetime and are associated to a well-defined notion of energy. In quantum gravity, while the specifics of any renormalization scheme will depend on the approach being considered, we should expect only a more abstract notion of scale, more or less reduced to a counting of degrees of freedom, to be available. This does not mean that such notion of scale cannot be tentatively interpreted in some proto-spatiotemporal manner, but it means that such interpretation will be fully justified only in the regime of the theory where continuum spacetime and geometry are shown to emerge.

Computing the full renormalization group flow of a given theory, in fact, amounts (formally) to defining the full (quantum) dynamics of the same \cite{FRG}; this means removing (again, formally) any truncation that may initially be applied to it to have mathematically well-defined quantities, in particular any truncation to a finite (small) number of its fundamental degrees of freedom. In this sense, it contains both a continuum limit (usually associated with the limit of large momenta and small distances) and a thermodynamic limit (usually corresponding to large volumes and infinite number of atoms or field values), when the two notions both make sense but differ, as in usual spacetime physics. We do not know if this is the case in quantum gravity, in general, and the question can be addressed only by considering specific quantum gravity formalisms and providing a definition of both limits. We do not distinguish the two, in what follows, referring simply to the \lq continuum limit\rq . 

The renormalization group flow is always computed within some approximation scheme. This is a technical necessity, but it also contains an important physical insight: what matters at each step is to control the (approximate) behaviour of key observables, and only the aspects of the theory which are most relevant to them, neglecting the rest. The important insight is that a lot of what the theory contains, in principle, does not affect the relevant physics, at least not significantly. The relevant collective variables and observables may differ from those in the initial definition of the same theory (and that enter the computation of the same collective quantities). Such approximate, coarse grained description, alongside other forms of truncation, is thus not just a technical tool that physicists have to adopt for lack of computational power or skills, but actually it is where they show or test their physical understanding of the system. These approximations, moreover, are both a pre-requisite for the understanding of the renormalization group flow of the theory and directly suggested by it, since the renormalization group flow itself gives indications on which dynamics and which observables (e.g. order parameters) are relevant in different regimes (scales). 

Next, we ask the second question: what should we expect to find, once we move along the $N$-direction via renormalization group tools, and as we approach the full quantum gravity corner of the Bronstein hypercube? 
A continuum spacetime and geometry is the goal, of course. And it is what we should find if our quantum gravity formalism is to be physically viable. But we should also expect to find much more than that. Physical quantum systems, when they are interacting (thus non-trivial) and possess an infinite (or very large) number of degrees of freedom, do not have a unique continuum limit. How they organize themselves when large numbers of their constituents are taken into account depends on the value of their fundamental coupling constants (or other external parameters). Their collective behaviour leads to different macroscopic phases, separated by phase transitions. The effective, emergent physics of different continuum phases can be very different, and they are also stable (by definition) under moderate changes of their defining parameters and, of course, under the dynamics of the system. They are in many ways different possible worlds, inequivalent collective realizations of what we were initially considering a single physical system, when focusing on small numbers of its fundamental constituents. The questions in any quantum gravity approach, once the fundamental entities and their quantum dynamics are identified, and the renormalization group flow can be, in principle, set rolling, become: what are the macroscopic phases? is any of them effectively described in terms of smooth geometry (with matter fields) and spacetime?
With these new questions come many others, for example concerning the physical meaning of the phase transition(s) separating a non-geometric from a geometric phase, that of the different geometric phases themselves, if more than one such phase appears, and the possible observational signatures of this potentially new physics.  

Before we survey briefly recent progress on the above issues, let us draw two general consequences of the above line of reasoning. First, moving along the $N$-direction, i.e. towards the full definition of the theory, brings potentially new physics and requires, possibly, a change in description of the system, at each step. Second, the result of such journey is not unique but it be given by different possible \lq continuum limits\rq, different continuum theories with different effective physics. In this sense, the Bronstein hypercube should not be expected to \lq close\rq ~to form an hypercube at all, but a multiplicity of possible hypercubes at best. And this is assuming that moving along the $h$ direction starting from the GR corner (or the one corresponding to some other continuum classical gravitational theory) gives any consistent result at all, and that, in addition, the $h$ and $N$ direction commute. This is also the point, however, where our attitude towards the non-spatiotemporal structures that our quantum gravity formalism is built on, whether we regard them as physical or mere mathematical (e.g. regularization) tools becomes crucial. In the latter case, in fact, only the continuum phase(s) with a geometric interpretation will be deemed physical, and the others ignored, and the result of quantizing the classical theory will have to be the same by definition, since the extension from the Bronstein cube to the hypercube would be a mere technical expedient, with the true physics remaining captured by the perspective associated with the Bronstein cube.

One more interesting and challenging aspect of the hypercube perspective on quantum gravity concerns symmetries, which are crucial tools in theory building \cite{symmetries}. Investigating the role of symmetries in 
quantum gravity from the point of view of the Bronstein hypercube means investigating two (related) issues: first, the fate of diffeomorphism invariance, the defining symmetry of General Relativity, in going from the Bronstein cube to the Bronstein hypercube, and then within the latter; second, the nature of new \lq quantum\rq ~symmetries and their use in moving along the $N$ direction and in recovering continuum spacetime and geometry in the end. Diffeomorphism invariance stands and falls, in many ways, alongside continuum spacetime and geometry, whether one sees it as a mere mathematical convenience or as a fundamental requirement of physical theories \cite{diffeomorphisms}. This means that, in the Bronstein cube, it will follow geometry and the gravitational field in being \lq quantized\rq, to reach the quantum gravity corner. This is nowhere more clear than in the canonical approach, where defining and imposing diffeomorphism invariance at the quantum level is equivalent to defining and imposing the quantum gravitational dynamics. It also means that, when changing the nature of the fundamental degrees of freedom of the theory, thus moving to candidate fundamental entities different than spacetime fields defined on some differentiable manifold, diffeomorphisms will cease to be defined altogether. Formally, at least, they will not play no role in the fundamental theory. This does not mean, however, that diffeomorphism symmetry will not be still important, at the conceptual level. To start with, the fundamental dynamics will have to be some form of constrained or relational dynamics, like in continuum, diffeomoprhism invariant GR. This is simply due to the fact that, like continuum diffeomorphism invariant GR, the fundamental quantum gravity theory will not admit any notion of time that is not given, at best, in terms of some internal (and approximate) degree of freedom of the theory, used as a relational clock, and thus will not take the form of a standard time evolution. This implies that the experience and mindset developed in the context of diffeomorphism invariant, continuum quantum theories will be very much relevant. Further, depending on the nature of the candidate fundamental degrees of freedom of the theory, the same may be characterised by new quantum symmetries which are some sort of pre-geometric counterpart of diffeomorphisms. They could be some discretised version of them, like in approaches based on simplicial piecewise-flat structures \cite{diffeoQG}, or of more exotic type, e.g. purely combinatorial \cite{CDT, CS}. In any case, the requirement that they reduce to diffeomorphisms in the appropriate continuum (and classical) limit, or that, at least, they can be traded for diffeomorphisms in the same limit, will play a crucial role in identifying such new symmetries. In turn, the identification of new fundamental symmetries, whatever their relation to diffeomorphisms, will certainly play an important role in the construction of the fundamental theory and in the reconstruction of continuum spacetime and geometry from it. They could be the key instrument for defining the fundamental theory space within which the dynamics of the pre-geometric atoms of space should be studied; they could be the tool for characterizing the different universality classes or continuum phases in which they organise themselves when moving along the $N$ direction; they could single out conserved quantities or other collective observables around which the emergent continuum geometric dynamics could be built. Work on all these aspects can be found in the literature, for example in GFT and random tensor models \cite{TheorySpace, GFTsymmetries, GFTphases, BFvacua}.

\section{A brief survey of recent results along the $N$ direction}
The renormalization of quantum gravity models is a very active area of research, in many of the quantum gravity formalisms based on non-spatiotemporal, discrete building blocks.  

It has been a central research topic in simplicial quantum gravity approaches since their very inception, also because in these approaches the discrete structures are usually seen as unphysical regularization tools so that the continuum limit is mandatory before thinking of any physics. In quantum Regge calculus \cite{qRC} one fixes the triangulation at the onset and has its edge lengths as dynamical variables, weighted by the (exponential of the)  Regge action (a discretization of the Einstein-Hilbert action for GR). The strategy for taking the continuum limit, usually limited to Euclidean geometries, is then analogous to the lattice gauge theory one, adapted to a varying lattice geometry, in which lattices are progressively refined by increasing their complexity while keeping some macroscopic quantity fixed, e.g. the total \lq spacetime volume\rq. There is no consensus on whether a continuum phase with a smooth geometry is found, and some of the phases identified so far are consistently interpreted as rather degenerate or singular from the continuum spacetime perspective, on the basis of simple indicators like dimension estimators (e.g. spectral or Hausdorff dimension). In dynamical triangulations one takes a complementary approach and, with the same starting point, it restricts attention to equilateral triangulations, which are summed over with the same Regge weight. The continuum limit amounts to computing the full sum over triangulations concentrating it on the finer ones (sending the fixed edge length to zero while keeping the total volume fixed). The results are consistent (and inconclusive) in the Euclidean setting, but become much more interesting when Lorentzian discrete geometries are considered and additional \lq causality\rq ~conditions are imposed, basically providing the triangulations summed over with a fixed foliation structure. Then, one finds strong indications that (at least) one smooth geometric phase is produced in the continuum limit, as it can be detected by dimension estimators and rather coarse geometric quantities, like the total spatial volume.

Random tensor models \cite{TM} produce in their perturbative expansion (around the fully degenerate configuration, corresponding to no spacetime at all) the same type of dynamical configurations, equilateral triangulations, and with the same weight, as the Euclidean dynamical triangulations approach. They provide a generating functional for them and, accordingly, offer a new set of statistical and field-theoretical tools to study their continuum limit. The results are so far broadly consistent with what has been found from the pure simplicial gravity perspective, in the simple large-$N$ limit of the tensors. The new set of tools, however, promises more and it has already hinted at a deeper level of analysis, with many preliminary results on double scaling limits, phases beyond what has been found in the large-$N$, etc. 

In canonical loop quantum gravity, maybe due to its traditional understanding as a straightforward quantization of continuum GR, the issue of renormalization and continuum limit has received less attention. A canonical renormalization group scheme at the full dynamical level has been proposed \cite{LQGrenorm}, but most of the results have been so far limited to the kinematical setting, neglecting the quantum dynamics of the theory. These results include the continuum limit that is used to define the kinematical Hilbert space of the theory and that singles out the so-called Ashtekar-Lewandowski vacuum as the basis for constructing spin network excitations \cite{LQG}. Its construction is a non-trivial mathematical achievement, but its physical nature is dubious. From the continuum perspective, it is a state corresponding to a totally degenerate geometry and a highly fluctuating connection, thus far away from anything resembling our spacetime. This prompts to look for the sector of the theory corresponding to highly excited states over such vacuum, and encoding many of the fundamental spin network degrees of freedom. It also suggests that, when this is done, the relevant description of the same theory will be very different, and possibly will involve a phase transition to a new, more geometric phase. Lacking a full renormalization group analysis, the issue of possible new vacua/phases could be studied only at the kinematical level, but produced already very interesting results. New kinematical vacua with a non-degenerate (constant) geometry, but still with highly fluctuating connection, were constructed and analysed in some detail \cite{timhanno}, shown to define inequivalent representations (thus genuinely new phases) and to offer already a more sensible physical interpretation in terms of continuum spacetime geometries. They can be understood as a sort of condensate of spin networks excitations, thus resonating with earlier \cite{GFTcontinuum} and more recent \cite{UniverseCondensate} work in the group field theory (re-)formulation of spin network dynamics. Even more recently, a different, complementary type of new vacua have been constructed \cite{BFvacua}, corresponding instead to a fixed curvature and fluctuating geometry (triad/flux variables), and with excitations corresponding to curvature defects. The simplest such vacuum can be associated to a simple BF topological field theory and zero curvature, while vacua corresponding to non-zero constant curvature (a cosmological constant?) seem to be given by condensates of curvature defects.  
 
Spin foam models \cite{SF} can be understood as a covariant dynamics for canonical loop quantum gravity states, i.e. spin networks, thus they are an alternative setting to define the renormalization group flow and the continuum limit of the theory, encoding also the quantum dynamics. They can also be understood as lattice gravity path integrals in first order (tetrad+connection) variables, thus in direct relation with simplicial quantum gravity approaches. Renormalization of spin foam models has been tackled in two (complementary) frameworks. The first \cite{SFrenorm} treats them as analogous to lattice gauge theories on a fixed lattice, thus in line with the way the continuum limit of quantum Regge calculus is studied. However, it brings on board mathematical methods and insights of canonical loop quantum gravity and it also takes advantage of the direct resemblance with gauge theories. More recently, also key tools from quantum information theory and many-body systems, like tensor networks, have started to play an important role. Coarse graining steps and renormalization are encoded in maps between spin foam amplitudes associated to different scales, where the notion of scale here is tied to the combinatorial complexity of the underlying lattice, and they provide therefore a dynamical counterpart of the kinematical continuum limit that defines the Hilbert space of the canonical theory. The results are so far mostly confined to simplified models, rather than with the full-fledged amplitudes proposed for 4d quantum gravity, but they are already very interesting, and possibly indicative of more general lessons. For example, one finds hints of a non-trivial phase diagram with a degenerate geometric phase and a non-trivial phase of topological nature, thus tentatively supporting the kinematical results on possible vacua in the canonical theory. The other way to tackle spin foam renormalization is to see them as Feynman amplitudes of group field theory models and to focus on the renormalization of the latter.

Group field theory renormalization has been in fact a very active and rapidly growing research direction for almost ten years, now, with many results \cite{GFTrenorm}. The strategy is to rely on the close-to-standard field theoretic formulations of these models as field theories on Lie group manifolds (not interpreted, of course, as spacetimes) and use their intrinsic notion of scale as a \lq distance on the group manifold\rq ~or, conversely, the conjugate momentum/resolution scale. Indeed, on such premises one can apply standard renormalization group techniques, suitably adapted to the peculiar combinatorially non-local nature of the GFT interactions. The trivial Fock vacuum of the theory, around which one sees spin network excitations and develops the perturbative expansion of a given group field theory model, is again a fully degenerate one, with no topological nor proto-geometrical excitations, thus it is in the non-perturbative sector of the theory that one looks for continuum spacetime, geometry and physics. Perturbative renormalization is however where one can find a consistency check of the quantum theory, a way to constrain model-building ambiguities, and deal more directly with spin foam amplitudes. This activity relied heavily on the parallel results on random tensor models, in particular the large-$N$ expansion, and has been also focused on simplified models. It has produced rigorous proofs of renormalizability of a wide range of tensorial GFTs via multi-scale methods: abelian and non-abelian, with local gauge invariance, thus having Feynman amplitudes corresponding to lattice gauge theories (and spin foam models) and without it, in low (e.g. 3d) as well as higher (e.g. 6d) topological dimensions. More recently, work on non-perturbative renormalization has gained traction, with the development of the Functional Renormalisation Group formalism for GFTs. The same range of models studied perturbatively has been studied by the FRG method, and there is by now a large body of results establishing renormalizability as well as non-trivial phase diagrams for many tensorial GFTs, their asymptotic freedom or safety in the UV, but also solid hints of Wilson-Fisher type fixed points in the IR, suggesting the existence of a condensate phase in the continuum limit. This phase would be directly relevant for the extraction of effective spacetime physics, especially in the cosmological setting, as we discuss below. For full-blown GFT models for 4-dimensional spacetime, we have mainly partial results on radiative corrections \cite{EPRLdiverg}, but the mathematical technology at our disposal is improving fast, and we can rely also on related analyses of GFT models on homogeneous spaces, and on existence proofs of phase transitions in the GFT formulation of topological BF theories in any dimension (the basis of a lot of model building for 4d quantum gravity models) \cite{BFphase}.

While the renormalization analysis of quantum gravity models has grown in attention and results, comparatively little work has been done so far on the extraction of continuum physics from them. By this we intend the extraction of effective dynamics for collective observables with a spatiotemporal and geometric interpretation, in the regime in which large numbers (possibly infinite) of the fundamental constituents are accounted for, and thus employing a set of explicit approximations and coarse graining operated at the level of the fundamental theory itself. We do not refer here to the many works in which tentative physics is extracted from either simple models of continuum spacetime and gravitational dynamics that are only inspired, but not derived from the fundamental theory (e.g. loop quantum cosmology, and several models of quantum black holes), or truncations of the fundamental theory dealing with very small numbers of the fundamental entities (e.g. restricted to simple spin network graphs, or simple spin foam or simplicial gravity lattices). 

One example of this type of work is the extraction of an effective minisuperspace dynamics and of a de Sitter-like (spatial) volume profile from the causal dynamical triangulations approach, obtained by explicit numerical evaluation of this (very coarse-grained) observable \cite{CDT}.

Another example of recent work in this direction is  GFT condensate cosmology \cite{GFTcosmo}, based on two main assumptions, one of perspective, one more technical. The first is that cosmology has to be looked for in the hydrodynamics of the fundamental theory, as the most suitable approximation for close-to-equilibrium and most coarse-grained dynamics, in the continuum limit. The second is that the relevant class of continuum states (implicitly, the  relevant continuum phase of the theory) for the extraction of gravitational physics is that captured by condensates of the microscopic building blocks (GFT quanta, i.e. spin network vertices or basic simplices). The first assumption suggests concepts and technical tools to be used. The second, supported also by the hints coming from the renormalization analysis of GFT models, makes it possible to go directly from the microscopic definition of the quantum dynamics of any given model, including the more promising 4d gravity ones, to an effective dynamics for cosmological observables in the continuum limit. For quantum condensates, in fact, the continuum hydrodynamics corresponds, in the simplest (mean field, Gross-Pitaevskii) approximation, to the classical equations of motion of the underlying field theory for the \lq atoms\rq, and the same happens in GFT models. This strategy led to many results, over the last five years. They include: the extraction of a modified Friedmann dynamics for homogeneous and isotropic geometries (and scalar matter), whose physics is captured by relational observables\footnote{Let us add some clarifications about the  use of relational (dynamical) variables in a pre-geometric, non-spatiotemporal quantum gravity formalism. The use of relational clocks and rods is a convenient strategy for defining a diffeomorphism invariant and physical notion of time and space, often adopted in General Relativity and in its quantized counterparts. We find it a perfectly acceptable solution to such issue (even if such clocks and rods will not be idealised or perfect, since they remain physical and quantum entities, and thus will offer only an approximate substitute for temporal or spatial coordinates). It is a solution, that is, to the issue of \lq disappearance of space and time\rq , in the sense proper to (quantum) General Relativity of disappearance of any absolute, non dynamical notion of space and time (as codified by a preferred frame), due to diffeomorphism invariance and background independence \cite{Carlo}. In the relational strategy, instead of any such absolute space or time, we have dynamical fields, which may include the metric field itself, playing the role of clocks and rods. This strategy requires, for its adoption, the usual continuum setting of the Bronstein cube, in which dynamical fields are basic entities. The adoption of the same strategy in a pre-geometric, non-spatiotemporal setting as pictured in the Bronstein hypercube should be understood in the following generalised sense. One looks at the fundamental degrees of freedom of the quantum gravity formalism at hand, which are non-spatiotemporal by definition, and do not correspond directly to continuum fields. Among those, one identifies the ones that, in a continuum limit and in an approximate sense (that is, after moving along the $N$ direction towards the standard continuum, spatiotemporal setting), will correspond to emergent dynamical fields, that can then be used as relational clock and rods, i.e. in terms of which one will then assign spatiotemporal localisation properties to the {\it other} degrees of freedom of the quantum gravity system. Such use as relational notions of space and time will take place, in other words,  only after a spatiotemporal GR-like description has emerged (or in order to test its emergence). It is possible that the application of this strategy requires the introduction of additional degrees of freedom in a given model of pre-geometric model, but it does not affect the need to \lq move along the $N$ direction\rq, for the whole set of degrees of freedom, to show the emergence of spacetime along the way (thus it does not mean that one is introducing space or time \lq by hand\rq in the fundamental definition of the theory, just as the fact that eventually spacetime and geometry emerge approximately in a pre-geometric quantum gravity model does not imply that they were there from the very beginning).}, with the correct classical limit at late times; a quantum bounce replacing the classical big bang singularity, as long as one remains within the hydrodynamics approximation of the full theory; the possibility of a long-lasting accelerated phase of expansion after such bounce (a sort of purely quantum gravity induced inflation) without the need for introducing any inflaton-like field; some preliminary study of the dynamics of anisotropies, showing their natural suppression as the universe grows; the first extensions of the formalism to cosmological perturbations, that seems to indicate how a scale invariant spectrum is the natural outcome of the dynamics, as long as one remains close to homogenous condensate states. To these, one could add the first generalization of the scheme to spherically symmetric geometries and black hole horizons \cite{GFTblackholes}, with more interesting results. We are just at the beginning of the exploration of the emergent continuum physics of GFT models (and of their spin foam counterpart), clearly, but the path seems promising.

\section*{Conclusions}
We have argued that the proper setting for thinking about quantum gravity, and for exploring the many issues it raises (mathematical, physical, conceptual), is broader than the traditional one of \lq quantizing GR\rq, well captured by the Bronstein cube. It is best pictured as a Bronstein hypercube, in which the non-spatiotemporal nature of the fundamental building blocks suggested by most quantum gravity formalisms (and even by semi-classical physics), and the need to control their collective dynamics, are manifest. This allows the proper focus on the problem of the emergence of continuum spacetime and geometry from such non-spatiotemporal entities. We have also argued that modern quantum gravity approaches are well embedded into this conceptual scheme, and have already started producing many results on the issues that are put to the forefront by it.
The quantum gravity world is therefore even richer, more complex but also more exciting than traditionally thought, and we are already actively exploring it. More surprises should be expected.

\end{document}